\documentclass[twocolumn,showkeys]{revtex4}
\usepackage[utf8]{inputenc}
\usepackage{graphicx}

\begin{document}

\title{Relativistic hydrodynamic model with the average reverse gamma factor evolution for the degenerate plasmas:
high-density ion-acoustic solitons}

\author{Pavel A. Andreev}
\email{andreevpa@physics.msu.ru}
\affiliation{Department of General Physics, Faculty of physics, Lomonosov Moscow State University, Moscow, Russian Federation, 119991.}

\date{\today}

\begin{abstract}
High-density low-temperature plasmas with degenerate species are considered
in the limit of high Fermi velocities close to the speed of light.
The small amplitude ion-acoustic solitons are studied in this regime.
Presented analysis is based on the relativistic
hydrodynamic model with the average reverse gamma factor evolution
consists of the equations for evolution of the following functions
the concentration, the velocity field,
the average reverse relativistic gamma factor, and the flux of the reverse relativistic gamma factor,
which are considered as main hydrodynamic variables.
Justification of the suggested model via comparison of the hydrodynamic results with the result of application of the relativistic Vlasov kinetic equation
is made in the linear approximation.
\end{abstract}

\keywords{relativistic plasmas, hydrodynamics, microscopic model, degenerate electrons.}


\maketitle





\section{Introduction}

Different fields of relativistic plasmas are under current development
due to the description of various astrophysical scenarios
and generation of relativistic plasmas in laboratories at the propagation of high-energy density electromagnetic beams
\cite{Shatashvili PoP 20}, \cite{Liu PPCF 21}, \cite{She 21}, \cite{Bhattacharjee PRD 19}, \cite{Comisso 19},
\cite{Darbha 21}, \cite{Chabanov 21}, \cite{Mahajan PoP 16}, \cite{Mahajan PoP 2011},
\cite{Comisso PRL 14}, \cite{Heyvaerts AA 12}, \cite{Mahajan PRL 03}, \cite{Munoz EPS 06}, \cite{Brunetti MNRAS 04}.

The degenerate plasmas can be placed in the middle between the classical and quantum plasmas.
It can be considered as the quasi-classic object described by the classic hydrodynamic or kinetics,
where the equation of state is chosen as the Fermi pressure or the equilibrium distribution function is chosen as the Fermi step
(the zero-temperature limit of the Fermi-Dirac distribution function).
However, the Fermi pressure and the Fermi step distribution function are caused by the Pauli blocking,
which is the quantum effect.
But, other quantum effects like the quantum Bohm potential are secondary in the degenerate plasmas for the majority of physical scenarios.

In this paper we focus on relativistic plasmas.
Relativistic regime can be reached within different scenarios.
Monoenergetic beams of electrons can propagate through the plasmas,
with the velocity of beam close to the speed of light.
There is the opposite limit of the macroscopically motionless plasmas heated up to temperatures of order of
the rest energy of at leats lightest of species of plasmas
(here and below we consider temperature in the energy units).
This regime is rather more complex in compare with monoenergetic beams
since it requires to get relation between the momentum density and the velocity field
for the large number of particles moving with different velocities
(if we use the Euler equation in the form of the momentum balance equation).
The degenerate relativistic plasmas is the low-temperature high-density plasmas,
where the Fermi velocity is of order of the speed of light.
So, we have distribution of particles over quantum states with different energies
and we have same complexity of the model like for the thermally distributed plasmas.

Novel model for the description of the relativistically hot plasmas is developed in Refs.
\cite{Andreev 2021 05}, \cite{Andreev 2021 06}, \cite{Andreev 2021 07}, \cite{Andreev 2021 08}, \cite{Andreev 2021 09},
\cite{Andreev 2021 10}.
This model is based on the derivation of hydrodynamic equations directly from their microscopic motion described by
the corresponding relativistic form of the second Newton's law as the equations of motion of each particle.
This method can be considered as the reformulation of the classical mechanics in the form of equations of field
since no probabilistic method is used during derivation.
Originally this method is suggested for the derivation of the kinetic Vlasov equation for the relativistic plasmas \cite{Kuz'menkov 91}.
Next, its simplification is presented for the hydrodynamic modeling of the nonrelativistic plasmas \cite{Drofa TMP 96}, \cite{Andreev PIERS 2012}.
Finally, an original structure of the hydrodynamic model for the relativistic plasmas is derived using this method
\cite{Andreev 2021 05}, \cite{Andreev 2021 06}, \cite{Andreev 2021 07}, \cite{Andreev 2021 08}, \cite{Andreev 2021 09},
\cite{Andreev 2021 10}.

The model itself consist of equations for evolution of four functions,
two three-scalars and two three-vectors,
which can be combined in two four-vectors.
These functions are the concentration of particles,
the velocity field obtained via the evolution of the current of particles,
the average reverse relativistic gamma-factor,
and
the current of the average reverse relativistic gamma-factor.
Basically, the average reverse relativistic gamma-factor is the ariphmetic average of $\sqrt{1-\textbf{v}_{i}^{2}(t)/c^{2}}$
over all particles,
where
$\textbf{v}_{i}(t)$ is the velocity of $i$-th particle as the function of time $t$ in accordance with the microscopic equation of motion,
$c$ is the speed of light.

A question might appear: Why do we use so strange functions like "the average reverse relativistic gamma-factor" in our model?
The answer is following.
We do not try to choose functions for the description of plasmas,
but we follow the structure of hydrodynamic equations as it appears during derivation.
The evolution of the concentration $n$ leads to the current of particles $\textbf{j}$.
Therefore, we consider the evolution of the current of particles $\textbf{j}$,
which has no direct relation to the momentum density in contrast with the nonrelativistic regime,
where these functions coincide.
The current of particles $\textbf{j}$ allows us to introduce the velocity field $\textbf{v}\equiv \textbf{j}/n$.
Hence, the equation for evolution of the current of particles $\textbf{j}$ gives the equation for evolution of the velocity field $\textbf{v}$.
This equation contains four novel functions.
One is the second rank tensor describing the flux of current of particles.
Three other functions appear at the presentation of interaction in the self-consistent (mean-field) approximation.
Two of them are mentioned above:
the average reverse relativistic gamma-factor,
and
the current of the average reverse relativistic gamma-factor.
Another one is the second rank tensor of flux of the current of the average reverse relativistic gamma-factor.
Consequently, we use
the average reverse relativistic gamma-factor,
and
the current of the average reverse relativistic gamma-factor
in order to continue the set of hydrodynamic equations and find appropriate regime of truncation.
The derivation shows that
the evolution of
the average reverse relativistic gamma-factor,
and
the current of the average reverse relativistic gamma-factor mostly
leads to reappearance of the concentration and the current of particles.
Hence, on this stage the mathrematical structure of the model shows some tendency to close itself relatively presented set of functions.
Obviously, there is no complete closer of the set of equation
with no additional truncation due to the large number of degrees of freedom in the many-particle systems.

Truncation is made via applications of equations of state for the second and higher rank tensors.
In this paper we consider the degenerate electron gas.
Therefore, we derive corresponding equations of state for the zero temperature regime describe by the Fermi step distribution function.


This paper is organized as follows.
In Sec. II the relativistic hydrodynamic model is adopted for the degenerate electron gas.
In Sec. III the linear and nonlinear analysis of the small amplitude longitudinal waves is given.
The accuracy of the suggested model is also analyzed via comparison of the Langmuir wave spectrum with the result obtain in the kinetic model.
The ion-acoustic solitons are considered by the reductive perturbation method.
In Sec. IV a brief summary of obtained results is presented.


\section{Relativistic hydrodynamic model}

Quasiclassic analysis of the degenerate electron gas can be based on the classical hydrodynamic model,
where the equations of state are obtained within the distribution function in form of the Fermi step
(the zero temperature limit of the Fermi-Dirac distribution function).

The nonrelativistic classic hydrodynamics obtained in the selfconsistent field approximation requires some equation of state for the pressure $P^{ab}$.
The relativistic hydrodynamic model based on the momentum balance equation requires two equations of state.
One equation of state is for the pressure.
The second equation of state is necessary for the momentum density
in order to make transition to the velocity field
which exists in the continuity and Maxwell equations.
The relativistic hydrodynamic model with the average reverse gamma factor evolution considered in this paper includes
evolution of the concentration, the velocity field,
the average reverse relativistic gamma factor, and the flux of the reverse relativistic gamma factor.
Therefore, it requires three equations of state.
Two equations of state are for the second rank tensors describing
the flux of the current of particles $p^{ab}$
and
the current of flux of the reverse relativistic gamma factor.
One equation of state is for the fourth rank tensor $M^{abcd}$,
which is the flux of the current of tensor $p^{ab}$.

Here we follow Refs. \cite{Andreev 2021 05}, \cite{Andreev 2021 09},
and adopt the model for the relativistic degenerate plasmas.
Hence, we have the Fermi velocity $v_{Fe}$ comparable with the speed of light $c$.

First equation in the presented model is the continuity equation \cite{Andreev 2021 05}
\begin{equation}\label{RHD2021ClLM cont via v} \partial_{t}n+\nabla\cdot(n\textbf{v})=0.\end{equation}

Next, the velocity field evolution equation is \cite{Andreev 2021 05}, \cite{Andreev 2021 09}
$$n\partial_{t}v^{a}+n(\textbf{v}\cdot\nabla)v^{a}
+\frac{1}{m}\partial^{a}\tilde{p}$$
$$=\frac{e}{m}\biggl(\Gamma -\frac{\tilde{t}}{c^{2}}\biggr)E^{a}+\frac{e}{mc}\varepsilon^{abc}(\Gamma v_{b}+t_{b})B_{c}$$
\begin{equation}\label{RHD2021ClLM Euler for v}
-\frac{e}{mc^{2}}(\Gamma v^{a} v^{b}+v^{a}t^{b}+v^{b}t^{a})E_{b}, \end{equation}
where tensor
$p^{ab}=\tilde{p}\delta^{ab}$
is the flux of the thermal velocities,
and tensor $t^{ab}=\tilde{t}\delta^{ab}$
is the flux of the average reverse gamma-factor.
Parameter $\tilde{p}$ is used here instead of $p$ presented in earlier papers \cite{Andreev 2021 05} in order to distinguish it from the momentum.
Parameters $m$ and $e$ are the mass and charge of particle,
$c$ is the speed of light,
$\delta^{ab}$ is the three-dimensional Kronecker symbol,
$\varepsilon^{abc}$ is the three-dimensional Levi-Civita symbol.
In equation (\ref{RHD2021ClLM Euler for v}) and below we assume the summation on the repeating index
$v^{b}_{s}E_{b}=\sum_{b=x,y,z}v^{b}_{s}E_{b}$.
Moreover, the metric tensor has diagonal form corresponding to the Minkovskii space,
it has the following sings $g^{\alpha\beta}=\{-1, +1, +1, +1\}$.
Hence, we can change covariant and contrvariant indexes for the three-vector indexes: $v^{b}_{s}=v_{b,s}$.
The Latin indexes like $a$, $b$, $c$ etc describe the three-vectors,
while the Greek indexes are deposited for the four-vector notations.
The Latin indexes can refer to the species $s=e$ for electrons or $s=i$ for ions.
The Latin indexes can refer to the number of particle $j$ at the microscopic description.
However, the indexes related to coordinates are chosen from the beginning of the alphabet,
while other indexes are chosen in accordance with their physical meaning.

The equation of evolution of the averaged reverse relativistic gamma factor includes the action of the electric field
$$\partial_{t}\Gamma+\partial_{b}(\Gamma v^{b}+t^{b})$$
\begin{equation}\label{RHD2021ClLM eq for Gamma}
=-\frac{e}{mc^{2}}n(\textbf{v}\cdot\textbf{E})
\biggl(1-\frac{1}{c^{2}}\biggl(\textbf{v}^{2}+\frac{5\tilde{p}}{n}\biggr)\biggr).\end{equation}
Function $\Gamma$ is also called the hydrodynamic Gamma function \cite{Andreev 2021 05}.

The fourth and final equation in this set of hydrodynamic equations is the equation of evolution for the thermal part of
current of the reverse relativistic gamma factor (the hydrodynamic Theta function):
$$(\partial_{t}+\textbf{v}\cdot\nabla)t^{a}
+\partial^{a}\tilde{t}
+(\textbf{t}\cdot\nabla) v^{a}+t^{a} (\nabla\cdot \textbf{v})
+\Gamma(\partial_{t}+\textbf{v}\cdot\nabla)v^{a}$$
$$
=\frac{e}{m}nE^{a}\biggl[1-\frac{\textbf{v}^{2}}{c^{2}}-\frac{3\tilde{p}}{nc^{2}}\biggr]
+\frac{e}{mc}\varepsilon^{abc}nv_{b}B_{c}
\biggl[1-\frac{\textbf{v}^{2}}{c^{2}}-\frac{5\tilde{p}}{nc^{2}}\biggr]$$
\begin{equation}\label{RHD2021ClLM eq for t a}
-\frac{2e}{mc^{2}}\Biggl[E^{a}\tilde{p}\biggl(1-\frac{\textbf{v}^{2}}{c^2}\biggr)
+nv^{a}v^{b}E_{b}\biggl(1-\frac{\textbf{v}^{2}}{c^{2}}-\frac{9\tilde{p}}{nc^{2}}\biggr)
-\frac{5M_{0}}{3c^{2}} E^{a}\Biggr].\end{equation}
All hydrodynamic equations are obtained in the mean-field approximation (the self-consistent field approximation).
The fourth rank tensor $M^{abcd}$ entering the equation for evolution of the flux of reverse gamma factor
via its partial trace $M^{abcc}=M_{c}^{cab}$.
In the isotropic limit,
we construct tensor $M^{abcd}$ of the Kronecker symbols
$M^{abcd}=(M_{0}/3)(\delta^{ab}\delta^{cd}+\delta^{ac}\delta^{bd}+\delta^{ad}\delta^{bc})$.
It gives $M^{xxxx}=M^{yyyy}=M^{zzzz}=M_{0}$.
If we have two pairs of different projections
we obtain $M^{xxyy}=M^{xxzz}=M_{0}/3$.
Otherwise the element of tensor $M^{abcd}$ is equal to zero.
So, for the partial trace $M_{c}^{cab}$ we find $M_{c}^{cab}=(5M_{0}/3)\delta^{ab}$.
For the explicit definition of tensor $M^{abcd}$ see equation (17) of Ref. \cite{Andreev 2021 05}.

The equations of electromagnetic field have the traditional form
presented in the three-dimensional notations
\begin{equation}\label{RHD2021ClLM div B} \nabla \cdot\textbf{B}=0,\end{equation}
\begin{equation}\label{RHD2021ClLM rot E} \nabla\times \textbf{E}=-\frac{1}{c}\partial_{t}\textbf{B},\end{equation}
\begin{equation}\label{RHD2021ClLM div E with time} \nabla \cdot\textbf{E}=4\pi(en_{i}-en_{e}),\end{equation}
and
\begin{equation}\label{RHD2021ClLM rot B with time}
\nabla\times \textbf{B}=\frac{1}{c}\partial_{t}\textbf{E}+\frac{4\pi q_{e}}{c}n_{e}\textbf{v}_{e},\end{equation}
where the ions exist as the motionless background.

\subsection{Equations of state}

Equations (\ref{RHD2021ClLM cont via v})-(\ref{RHD2021ClLM eq for t a}) originally obtained for the relativistically hot plasmas.
One of its features is the large temperature,
so the effective thermal velocities are close to the speed of light.
Let us mention that
the zero temperature limit of these equations can also be considered to study the propagation of the monoenergetic beams \cite{Andreev 2021 09}.
However, the opposite limit is under consideration in this paper.
As it is mentioned above
we are going to make a substitution in order to consider the degenerate electron gas of high concentration,
so the Fermi velocity $v_{Fe}=p_{Fe}/\sqrt{1+p_{Fe}^{2}/m^{2}c^{2}}m$,
where $p_{Fe}=(3\pi^{2}n)^{1/3}\hbar$.

In order to give a quasi-classic analysis of the high density relativistic degenerate electron gas
(or all species of plasmas)
we need to find corresponding equations of state for functions
$p^{ab}$, $t^{ab}$, and $M^{abcd}$.

Degenerate electrons are described within the zero-temperature limit of the Fermi-Dirac distribution,
which is given by the Fermi step distribution
\begin{equation}\label{RHD2021ClLM Fermi step} f_{0}=\Biggl\{\begin{array}{c}
                                                               \frac{2}{(2\pi\hbar)^{3}} \\
                                                               0
                                                             \end{array}
\textrm{for}
\begin{array}{c}
                                                               p\leq p_{Fe} \\
                                                               p> p_{Fe}
                                                             \end{array}
\end{equation}

Before we consider novel functions 
$p^{ab}$, $t^{ab}$, and $M^{abcd}$
we present the equation of state for the pressure as a point of reference.
The concentration has well-known form in terms of the distribution function
\begin{equation}\label{RHD2021ClLM concentr via f} n=\int f_{0} d^{3}p. \end{equation}
The pressure (the flux of the momentum density) can be written in the following forms:
\begin{equation}\label{RHD2021ClLM Pressure def via f} P^{ab}=\int p^{a}v^{b}f_{0} d^{3}p=c\int p^{a}p^{b}f_{0} d^{3}p/p_{0}, \end{equation}
where $p_{0}=\gamma mc=mc/\sqrt{1-v^{2}/c^{2}}$,
$\textbf{p}=m\textbf{v}/\sqrt{1-\textbf{v}^{2}/c^{2}}$,
$\gamma=1/\sqrt{1-v^{2}/c^{2}}$,
and the second expression is shown in more symmetric form including the covariant element of volume in the momentum space $d^{3}p/p_{0}$.
Calculation leads to $P^{ab}=P \delta^{ab}$,
with
\begin{equation}\label{RHD2021ClLM Pressure rel eq of state}
P= \frac{m^{4}c^{5}}{24\pi^{2}\hbar^{3}}\biggl[\xi\sqrt{\xi^{2}+1}(2\xi^{2}-3)+3Arsinh\xi\biggr],
\end{equation}
where
$\xi\equiv p_{Fe}/mc$,
$Arsinh\xi=ln\mid \xi+\sqrt{\xi^{2}+1}\mid$,
and
$sinh(Arsinh\xi)=\xi$.

At this step we are ready to go further and calculate expressions for the novel functions.
The first of the is the flux of the current of particles,
which has the following representation in form of the distribution function
\begin{equation}\label{RHD2021ClLM p via f} p^{ab}=\int v^{a}v^{b} f_{0} d^{3}p. \end{equation}
Next, we obtain the equation of state $p^{ab}=\tilde{p} \delta^{ab}$ with
\begin{equation}\label{RHD2021ClLM p rel eq of state}
\tilde{p}=\frac{m^{3}c^{5}}{3\pi^{2}\hbar^{3}}\biggl[\frac{1}{3}\xi^{3}-\xi+\arctan\xi\biggr],
\end{equation}
where $m^{3}c^{5}/3\pi^{2}\hbar^{3}=n_{0}c^{2}(m^{3}c^{3}/p_{Fe}^{3})$

The second of required functions is the flux of the current of the average reverse gamma factor
\begin{equation}\label{RHD2021ClLM t via f} t^{ab}=\int \biggl(\frac{v^{a}v^{b}}{\gamma}\biggr) f_{0} d^{3}p. \end{equation}
Our calculation leads to $t^{ab}=\tilde{t} \delta^{ab}$ with
\begin{equation}\label{RHD2021ClLM t rel eq of state}
\tilde{t}=\frac{m^{3}c^{5}}{6\pi^{2}\hbar^{3}} \biggl[ \xi\sqrt{\xi^{2}+1}+\frac{2\xi}{\sqrt{\xi^{2}+1}} -3Arsinh\xi\biggr].
\end{equation}

The fourth rank tensor should be also calculated using the Fermi step
\begin{equation}\label{RHD2021ClLM M via f} M^{abcd}=\int v^{a}v^{b} v^{c}v^{d} f_{0} d^{3}p. \end{equation}
It leads to the symmetric expression
$M^{abcd}=(M_{0}/3)(\delta^{ab}\delta^{cd}+\delta^{ac}\delta^{bd}+\delta^{ad}\delta^{bc})$
with
\begin{equation}\label{RHD2021ClLM M rel eq of state}
M_{0}=\frac{m^{3}c^{7}}{30\pi^{2}\hbar^{3}} \biggl[ 2\xi(\xi^{2} -6) -\frac{3\xi}{\xi^{2}+1} +15\arctan\xi \biggr].
\end{equation}

We also need to find the equilibrium expression for the average reverse gamma factor $\Gamma_{0}$ for the degenerate electron gas
\begin{equation}\label{RHD2021ClLM Gamma via f} \Gamma=\int \frac{1}{\gamma} f_{0} d^{3}p. \end{equation}
After calculation for the degenerate electron gas we obtain
\begin{equation}\label{RHD2021ClLM Gamma rel eq of state}
\Gamma= \frac{m^{3}c^{3}}{2\pi^{2}\hbar^{3}} \biggl[ \xi\sqrt{\xi^{2}+1} -Arsinh\xi\biggr].
\end{equation}
In this model the relativistic Gamma function $\Gamma$ is an independent function.
Its evolution is described by equation (\ref{RHD2021ClLM eq for Gamma}).
Equation (\ref{RHD2021ClLM Gamma rel eq of state}) is used as the equation of state for the equilibrium value of the relativistic Gamma function $\Gamma$.


\section{Waves in the relativistic magnetized plasmas}

\subsection{Equilibrium state and the linearized hydrodynamic equations}

We focus on degenerate electron-ion plasmas,
where both components are degenerate.
Moreover, the concentration of both components $n_{0e}=n_{0i}$ is high,
up to values giving large Fermi velocities $v_{Fe}$ and $v_{Fi}$ getting close to the speed of light $c$.

We consider small perturbations of the equilibrium state
while the equilibrium state is characterized by the constant concentrations of electrons and ions $n_{0e}$, $n_{0i}$,
constant values of the average reverse relativistic gamma factors $\Gamma_{0e}$, $\Gamma_{0i}$,
zero values of the equilibrium velocity fields of both species,
zero values of the current of average reverse relativistic gamma factors,
and zero values of the electric and magnetic fields.
So, the hydrodynamic functions are
$n_{s}=n_{0s}+\delta n_{s}$,
$v_{xs}=\delta v_{xs}$,
$\Gamma_{s}=\Gamma_{0s}+\delta \Gamma_{s}$,
$t_{xs}=\delta t_{xs}$,
$E_{x}=\delta E_{x}$,
perturbations of the magnetic field are not considered since we consider the longitudinal waves.
It is also assumed that the perturbations have the monochromatic form,
for instance $\delta n_{s}=N_{s}e^{-\imath\omega t+\imath k x}$,
where $N_{s}$ is the amplitude.

The linearized continuity equation has well-known form for the macroscopically motionless fluids
(no equilibrium velocity field)
\begin{equation}\label{RHD2021ClLM continuity equation lin 1D}
\partial_{t}\delta n_{s}+n_{0s}\partial_{x} \delta v_{xs}=0. \end{equation}

The second equation appears from equation (\ref{RHD2021ClLM Euler for v}) in the following form
\begin{equation}\label{RHD2021ClLM velocity field evolution equation lin 1D}
n_{0s}\partial_{t}\delta v_{xs}+\frac{\delta \tilde{p}_{0s}}{\delta n_{0s}}\partial_{x}n_{s}
=\frac{q_{s}}{m_{s}}\Gamma_{0s} \delta E_{x}-\frac{q_{s}}{m_{s}c^{2}}\tilde{t}_{0s}\delta E_{x},
\end{equation}
where parameters $\delta \tilde{p}_{0s}/\delta n_{0s}$ and $\tilde{t}_{0s}$ appear from equations of state presented above.

We obtain the linearized equations for $\delta\Gamma$ and $\delta t_{x}$ from
equations (\ref{RHD2021ClLM eq for Gamma})
and (\ref{RHD2021ClLM eq for t a})
in the following form
\begin{equation}\label{RHD2021ClLM evolution of Gamma lin 1D}
\partial_{t}\delta\Gamma_{s} +\Gamma_{0s}\partial_{x}\delta v_{xs}+\partial_{x}\delta t_{xs} =0, \end{equation}
and
$$\partial_{t}\delta t_{xs} +\partial_{x}\delta \tilde{t}_{s}-\frac{\Gamma_{0s}}{n_{0s}}\partial_{x}\delta \tilde{p}_{s}
+\frac{q_{s}}{m_{s}}\frac{\Gamma_{0s}^{2}}{n_{0s}}\delta E_{x}$$
\begin{equation}\label{RHD2021ClLM evolution of Theta lin 1D}
=\frac{q_{s}}{m_{s}}n_{0s}\delta E_{x} -\frac{5q_{s}}{m_{s}c^{2}}\tilde{p}_{0s}\delta E_{x} +\frac{10q_{s}}{3m_{s}c^{4}}M_{0s}\delta E_{x}, \end{equation}
where $M_{0s}^{xxcc}=(5/3)M_{0s}$.

The linearized Poisson equation has the well-known form
\begin{equation}\label{RHD2021ClLM Poisson equation lin}
\partial_{x}\delta E_{x}=4\pi (q_{e} \delta n_{e}+q_{i} \delta n_{i}). \end{equation}

Analysis of the set of linearized equations obtained at zero external fields shows that
it is enough to use
equations (\ref{RHD2021ClLM continuity equation lin 1D}),
(\ref{RHD2021ClLM velocity field evolution equation lin 1D}),
(\ref{RHD2021ClLM Poisson equation lin})
to get a closed set of equations for the longitudinal perturbations.

The linearized hydrodynamic equations (\ref{RHD2021ClLM continuity equation lin 1D})-(\ref{RHD2021ClLM Poisson equation lin})
contain $\Gamma_{0}$ and $\tilde{t}_{0}$,
but function $\tilde{p}$ enters these equations in two forms.
It appears as the equilibrium value of this function on the right-hand side of equation (\ref{RHD2021ClLM evolution of Theta lin 1D}).
However, the Euler equation (\ref{RHD2021ClLM velocity field evolution equation lin 1D}) contains the perturbation of function $\tilde{p}$:
$\delta \tilde{p}_{s}=(\delta p/\delta n)\delta n_{s}$,
so $u_{ps}^{2}\equiv (\delta p_{s}/\delta n_{s})$.
Therefore, we present the expression for $\delta p_{s}/\delta n_{s}$ obtained from equation (\ref{RHD2021ClLM p rel eq of state})
\begin{equation}\label{RHD2021ClLM  d p on  d n rel eq of state}
\frac{\delta\tilde{p}}{\delta n}=\frac{1}{3}c^{2}\frac{\xi^{2}}{\xi^{2}+1}.\end{equation}
It gives $v_{Fe}^{2}/3$ in the nonrelativistic limit.
While the ultrarelativistic limit leads to $c^{2}/3$.
Using the relativistic expression for the Fermi velocity we find that
$\delta\tilde{p}/\delta n=v_{Fe}^{2}/3$ in the ambient relativistic regimes.


\subsection{Spectrum of the Langmuir waves: Hydrodynamic description}

To give a simple illustration of the relativistic effects existing in the presented model
we consider one of fundamental wave effects in plasmas -- the Langmuir wave spectrum.
The relativistic Langmuir waves are considered in Ref. \cite{Andreev 2021 05} for the thermally distributed electrons with the relativistic temperatures.
Here we consider this problem for the degenerate electrons.
Equations
(\ref{RHD2021ClLM continuity equation lin 1D})-(\ref{RHD2021ClLM Poisson equation lin})
allow to find the following spectrum
\begin{equation}\label{RHD2021ClLM Langmuir wave H}
\omega^{2}=\biggl(\frac{\Gamma_{0}}{n_{0}}-\frac{u_{t}^{2}}{c^{2}}\biggr)\omega_{Le}^{2} +u_{p}^{2}k_{z}^{2}, \end{equation}
where parameters $\Gamma_{0}$, $u_{t}^{2}$, and $u_{p}^{2}$ should be obtained from equations
(\ref{RHD2021ClLM p rel eq of state})-(\ref{RHD2021ClLM Gamma rel eq of state}).
However, they have different appearance.
As it is mentioned above $u_{p}^{2}$ appears from the perturbation of pressure
(\ref{RHD2021ClLM p rel eq of state}) $p=p_{0}+\delta p$ with $\delta p=u_{p}^{2} \delta n$.
But parameter $u_{t}^{2}$ appears from the equilibrium value of function $u_{t}^{2}\equiv \tilde{t}_{0}/n_{0}$.
Necessary substitution leads to the following result
\begin{equation}\label{RHD2021ClLM Langmuir wave H2} \omega^{2}=\frac{\omega_{Le}^{2}}{\gamma_{Fe}}
+\frac{1}{3}c^{2}\frac{p_{Fe}^{2}}{p_{Fe}^{2}+m^{2}c^{2}}k_{z}^{2}, \end{equation}
where $\gamma_{Fe}=1/\sqrt{1-v_{Fe}^{2}/c^{2}}=\sqrt{1+p_{Fe}^{2}/m^{2}c^{2}}$
is the standard relativistic gamma factor considered for the Fermi velocity.
Expression (\ref{RHD2021ClLM Langmuir wave H2}) can be represented via the Fermi velocity as follows
$\omega^{2}=\frac{\omega_{Le}^{2}}{\gamma_{Fe}}
+\frac{1}{3}v_{Fe}^{2}k_{z}^{2}$.

\subsection{Spectrum of the Langmuir waves in the relativistic kinetics for the degenerate electrons}

It would be useful to give an analysis of accuracy of the hydrodynamic model presented within equations
(\ref{RHD2021ClLM cont via v})-(\ref{RHD2021ClLM eq for t a}).
To this end we compare the spectrum of the Langmuir waves obtained above
(\ref{RHD2021ClLM Langmuir wave H2})
with the result of the relativistic kinetics.
Therefore, we present the Vlasov kinetic equation
\begin{equation}\label{RHD2021ClLM Vlasov eq} \partial_{t}f_{e}+\textbf{v}\cdot\nabla f_{e}
+q_{e}\biggl(\textbf{E}+\frac{1}{c}\textbf{v}\times\textbf{B}\biggr)\cdot\frac{\partial f_{e}}{\partial \textbf{p}}=0,\end{equation}
with the corresponding form of the Poisson equation
\begin{equation}\label{RHD2021ClLM div E kin}
\nabla\cdot \textbf{E}=4\pi q_{e}\int f_{e}(\textbf{r},\textbf{p},t)d\textbf{p} +4\pi q_{i}n_{0i},\end{equation}
for the analysis of the Longitudinal waves.

Equilibrium distribution function is given be equation (\ref{RHD2021ClLM Fermi step}).
Here we consider the small perturbations of this distribution $f_{e}=f_{0}+\delta f$
with $\delta f=F e^{-\imath\omega t +i \textbf{k} \textbf{r}}$.
Moreover, let us consider the small perturbations for the waves propagating in the plasmas being in the external uniform magnetic field,
so we have $\textbf{E}=0+\delta \textbf{E}$ and $\textbf{B}=B_{0}\textbf{e}_{z}+\delta \textbf{B}$.
Therefore, the Vlasov kinetic equation (\ref{RHD2021ClLM Vlasov eq}) transforms to the following form for the linear approximation
\begin{equation}\label{RHD2021ClLM Vlasov eq lin} -\imath(\omega- \textbf{k}\cdot\textbf{v}) \delta f
-m\frac{v_{\perp}}{p_{\perp}}\Omega_{e}\frac{\partial \delta f}{\partial \varphi_{p}}
+q_{e}\delta\textbf{E}\cdot\frac{\partial f_{0}}{\partial \textbf{p}}=0,\end{equation}
where we use
$\Omega_{e}=q_{e}B_{0}/m_{e}c$,
$[\textbf{v}\times\delta \textbf{B}]\cdot(\partial f_{0}(p)/\partial \textbf{p})=0$,
$\textbf{v}=v_{\perp}\textbf{e}_{\perp}+v_{z}\textbf{e}_{z}$,
and $v_{\perp}/p_{\perp}=\sqrt{1-v^{2}/c^{2}}=1/\gamma$.

Equation (\ref{RHD2021ClLM Vlasov eq lin}) leads to the following solution
$$\delta f=\int_{c}^{\varphi}d\varphi'
\biggl[\biggl(\frac{q\gamma}{\Omega_{e}}\delta \textbf{E}\cdot\frac{\partial f_{0}}{\partial \textbf{p}}\biggr)_{\varphi'}\cdot $$
\begin{equation}\label{RHD2021ClLM delta f solution 1}
\cdot\exp\biggl(\frac{\imath\gamma}{\Omega_{e}}\int_{\varphi}^{\varphi'}d\varphi'' (\omega-\textbf{k}\cdot\textbf{v})_{\varphi''}\biggr)\biggr].
\end{equation}
Integration leads to
$$\delta f=\frac{q_{e}\gamma}{\Omega_{e}}\frac{1}{p}\frac{\partial f_{0}}{\partial p}
\exp\biggl(-\imath\frac{(\omega-k_{z}v_{z})\varphi -k_{x}v_{\perp}\sin\varphi}{\Omega_{e}/\gamma}\biggr)\cdot$$
\begin{equation}\label{RHD2021ClLM delta f solution 2}
\cdot\int_{c}^{\varphi}d\varphi' (\delta \textbf{E}\cdot \textbf{p})_{\varphi'}
\exp\biggl(\imath\frac{(\omega-k_{z}v_{z})\varphi' -k_{x}v_{\perp}\sin\varphi'}{\Omega_{e}/\gamma}\biggr).
\end{equation}

For the longitudinal waves $\delta \textbf{E}\parallel \textbf{k}$
propagating parallel to the external magnetic field $\textbf{k}\parallel \textbf{B}_{0}$
we obtain
$$\delta f=\frac{q_{e}}{\Omega_{e}\sqrt{1-\frac{v^{2}}{c^{2}}}}\frac{1}{p}\frac{\partial f_{0}}{\partial p} p_{z} \delta E_{z}\cdot$$
\begin{equation}\label{RHD2021ClLM delta f solution 3}
\cdot\exp\biggl(-\imath\frac{(\omega-k_{z}v_{z})\varphi}{\Omega_{e}\sqrt{1-\frac{v^{2}}{c^{2}}}}\biggr)
\int_{c}^{\varphi}d\varphi' \exp\biggl(\imath\frac{(\omega-k_{z}v_{z})\varphi'}{\Omega_{e}\sqrt{1-\frac{v^{2}}{c^{2}}}}\biggr).
\end{equation}

Final integration leads to the following expression for the perturbation of the distribution function
\begin{equation}\label{RHD2021ClLM delta f solution 4}
\delta f=q_{e}\frac{p_{z}}{p}\frac{\partial f_{0}}{\partial p} \frac{\delta E_{z}}{\imath(\omega-k_{z}v_{z})},
\end{equation}
where all relativistic effects are placed in $(\omega-k_{z}v_{z})$ via $v_{z}=p_{z}/m\gamma$.

Let us use solution (\ref{RHD2021ClLM delta f solution 4}) for the calculation of the perturbations of the concentration
\begin{equation}\label{RHD2021ClLM delta n via delta f}
\delta n= \int d\textbf{p} \delta f(\textbf{r}, \textbf{p},t)
\end{equation}
and substitute it in the Poisson equation to get the dispersion equation
\begin{equation}\label{RHD2021ClLM Disp eq from kin}
1+3\gamma_{Fe}\frac{\omega_{Le}^{2}}{p_{Fe}^{2}k_{z}^{2}/m^{2}}
\biggl[1-\frac{m\gamma_{Fe}\omega}{p_{Fe}k_{z}}\ln\biggl(\frac{m\gamma_{Fe}\omega+p_{Fe}k_{z}}{m\gamma_{Fe}\omega-p_{Fe}k_{z}}\biggr)\biggr]=0.
\end{equation}
In the limit $\omega\gg k_{z}v_{Fe}=p_{Fe}k_{z}/m\gamma_{Fe}$ dispersion equation gives
the following spectrum of the relativistic Langmuir waves in the degenerate electron gas
\begin{equation}\label{RHD2021ClLM Langm spectrum from kin}
\omega^{2}=\frac{\omega^{2}_{Le}}{\gamma_{Fe}} +\frac{3}{5}\frac{p_{Fe}^{2}k_{z}^{2}}{m^{2}\gamma_{Fe}^{2}}.
\end{equation}
It can be also represented in the following form
$\omega^{2}=\frac{\omega^{2}_{Le}}{\gamma_{Fe}} +\frac{3}{5}v_{Fe}^{2}k_{z}^{2}$.
Comparison with the results of hydrodynamic model presented above shows agreement up to the coefficient in front of $v_{Fe}^{2}$.
Which is the well-known difference existing in the nonrelativistic limit as well.
There is systematic way of solving of this problem via the extension of set of hydrodynamic equations to get complete agreement with the kinetic description
(see for instance \cite{Tokatly PRB 99}, \cite{Tokatly PRB 00}, \cite{Andreev JPP 21}).

Presence of the relativistic gamma factor taken for the Fermi velocity $\gamma_{Fe}$
in the denumerator in the first term of equation (\ref{RHD2021ClLM Langm spectrum from kin})
shows the decrease of the cut-off frequency of the Langmuir waves.
Moreover, presence of the square of relativistic gamma factor
in the denumerator in the second term of equation (\ref{RHD2021ClLM Langm spectrum from kin})
shows stronger decrease of the change of frequency at the grough of the wave vector.
Which reveals in the decrease of the group velocity as well.

\subsection{Spectrum of the ion-acoustic waves}

From equations
(\ref{RHD2021ClLM continuity equation lin 1D})-(\ref{RHD2021ClLM Poisson equation lin})
we find the following dispersion equation for the longitudinal waves in the electron-ion plasmas in order to obtain the spectrum of the low frequency ion-acoustic waves
\begin{equation}\label{RHD2021ClLM Disp eq e-i}
1=\biggl(\frac{\Gamma_{0e}}{n_{0e}}-\frac{u_{te}^{2}}{c^{2}}\biggr)\frac{\omega_{Le}^{2}}{\omega^{2}-u_{pe}^{2}k_{z}^{2}}
+\biggl(\frac{\Gamma_{0i}}{n_{0i}}-\frac{u_{ti}^{2}}{c^{2}}\biggr)\frac{\omega_{Li}^{2}}{\omega^{2}-u_{pi}^{2}k_{z}^{2}}. \end{equation}
As it is demonstrated above we can simplify the coefficients in front of the Langmuir frequencies expressing it via the relativistic gamma factor.
Moreover, the characteristic velocities $u_{pe}$ and $u_{pi}$ have simple expressions via the Fermi velocities
$u_{pe}^{2}=v_{Fe}^{2}/3$ and $u_{pi}^{2}=v_{Fi}^{2}/3$.
Therefore, we give representation of dispersion equation (\ref{RHD2021ClLM Disp eq e-i}) in simplified form
\begin{equation}\label{RHD2021ClLM Disp eq e-i}
1=\frac{\omega_{Le}^{2}}{\gamma_{Fe}(\omega^{2}-v_{Fe}^{2}k_{z}^{2}/3)}
+\frac{\omega_{Li}^{2}}{\gamma_{Fi}(\omega^{2}-v_{Fi}^{2}k_{z}^{2}/3)}. \end{equation}
For the frequencies being in interval $v_{Fe}^{2}k_{z}^{2}/3\gg\omega^{2}\gg v_{Fi}^{2}k_{z}^{2}/3$
equation (\ref{RHD2021ClLM Disp eq e-i}) gives the following spectrum of the ion-acoustic waves
\begin{equation}\label{RHD2021ClLM spectrum iaw}
\omega^{2}=\frac{\omega_{Li}^{2}}{\gamma_{Fi}(1+\frac{\omega_{Le}^{2}}{\gamma_{Fe}v_{Fe}^{2}k_{z}^{2}/3})},
\end{equation}
or, in the long wavelength  limit,
\begin{equation}\label{RHD2021ClLM spectrum iaw small k}
\omega^{2}
=\frac{m_{e}\gamma_{Fe}}{m_{i}\gamma_{Fi}}v_{Fe}^{2}k_{z}^{2}/3
=\frac{p_{Fe}^{2}k_{z}^{2}}{3m_{e}m_{i}\gamma_{Fe}\gamma_{Fi}}.
\end{equation}
The expression of frequency square in terms of the Fermi velocity contains additional factor equal to ratio of the gamma factors for electrons and ions
$\gamma_{Fe}/\gamma_{Fi}$.
However, the momentum has expression via the concentration equals to the nonrelativistic expression
$p_{Fe}=(3\pi^{2}n_{0e})^{1/3}\hbar$.
Hence, the second expression in equation (\ref{RHD2021ClLM spectrum iaw small k}) gives more clear physical picture,
where the frequency square is proportional to product of reverse gamma factors $(\gamma_{Fe}\gamma_{Fi})^{-1}$.

The minimal frequency square of the Langmuir wave is decreased by factor $\gamma_{Fe}^{-1}$
(see equation (\ref{RHD2021ClLM Langm spectrum from kin})).
While the frequency square of the ion-acoustic waves shows stronger decrease
since it contains $(\gamma_{Fe}\gamma_{Fi})^{-1}$,
where additional factor $\gamma_{Fi}^{-1}<1$ is included.

\subsection{Small amplitude ion-acoustic soliton}

The ion-acoustic solitons are considered in the high-density low-temperature electron-ion plasmas.
They are studied in the limit of small amplitude of the soliton.
Therefore, we apply the reductive perturbation method (see for instance \cite{Andreev_Iqbal PoP 16}).
This method includes the expansion of hydrodynamic functions as the series on the small amplitude with some scaling presented by parameter $\varepsilon$.
In accordance with this method
we introduce the following couple of variables
with the necessary scaling
\begin{equation}\label{RHD2021ClLM def of xi}\begin{array}{cc}
\xi=\varepsilon^{\frac{1}{2}}(z-Vt), & \tau=\varepsilon^{\frac{3}{2}}t,  \end{array}                                                                      \end{equation}
where
parameter $\tau$ is proportional to the larger degree of small parameter $\varepsilon$.
The parameter $\tau$ is called the slower time,
while faster dependence on time $t$ is included in parameter $\xi$.



we introduce an expansion of the hydrodynamic parameters on a small parameter $\varepsilon$
\begin{equation}\label{RHD2021ClLM expansion of n s}  n_{s}=n_{0s}+\varepsilon n_{1s}+\varepsilon^{2} n_{2s},\end{equation}
\begin{equation}\label{RHD2021ClLM expansion of v s}  v_{sz}=0+\varepsilon v_{1sz}+\varepsilon^{2} v_{2sz},\end{equation}
\begin{equation}\label{RHD2021ClLM expansion of Gamma}  \Gamma_{s}=\Gamma_{0s}+\varepsilon \Gamma_{1s}+\varepsilon^{2} \Gamma_{2s},\end{equation}
\begin{equation}\label{RHD2021ClLM expansion of t flux of G}  t_{sz}=0+\varepsilon t_{1sz}+\varepsilon^{2} t_{2sz},\end{equation}
and
\begin{equation}\label{RHD2021ClLM expansion of phi} \phi=0+\varepsilon \phi_{1}+\varepsilon^{3} \phi_{2},\end{equation}
where
function $\Gamma_{0s}$ is given by equation (\ref{RHD2021ClLM Gamma rel eq of state}),
and
$\phi$ is the potential of the electric field $\textbf{E}=-\nabla\phi$.

Equations of state (\ref{RHD2021ClLM Pressure rel eq of state})-(\ref{RHD2021ClLM M rel eq of state}) for functions $P$, $p$, $t$, and $M$ allows to get representations of these functions via the different combinations of $n_{0s}$, $n_{1s}$, $n_{1s}^{2}$, and $n_{2s}$.
We find the expressions presented below:
\begin{equation}\label{RHD2021ClLM expansion of p on epsilon}
\tilde{p}_{s}\approx \tilde{p}_{0s}+\varepsilon u_{ps}^{2} n_{1s}
+\varepsilon^{2} u_{ps}^{2} n_{2s} +\varepsilon^{2} \frac{v_{Fs}^{2}}{9\gamma_{Fs}^{2}n_{0s}} n_{1s}^{2},
\end{equation}
where $u_{ps}^{2}=v_{Fs}^{2}/3$,
and
\begin{equation}\label{RHD2021ClLM expansion of p on epsilon}
\tilde{t}_{s}\approx \tilde{t}_{0s}+\varepsilon \frac{v_{Fs}^{2}}{3\gamma_{Fs}} n_{1s},
\end{equation}
for function $M_{s0}$ we need equilibrium expressions only.


Presented method of expansion leads to the continuity equation considered in the first (lowest) order of the expansion and the second order of expansion
\begin{equation}\label{RHD2021ClLM cont eq expansion order 1} n_{0s}\partial_{\xi}v_{sx1}=U\partial_{\xi}n_{s1}, \end{equation}
and
\begin{equation}\label{RHD2021ClLM cont eq expansion order 2}
\partial_{\tau}n_{s1}-U\partial_{\xi}n_{s2}+\partial_{\xi}(n_{s0}v_{sz2}+n_{s1}v_{sz1})=0. \end{equation}
More accurately speaking we have coefficients $\varepsilon^{3/2}$ and $\varepsilon^{5/2}$ for the first and second orders of expansion, correspondingly.

We can integrate equation (\ref{RHD2021ClLM cont eq expansion order 1}) under boundary condition
that  the perturbation caused by soliton goes to zero at infinite distance from its center
$v_{sx1}\rightarrow0$ and $n_{s1}\rightarrow0$ at $\xi\rightarrow\pm\infty$.
We obtain
\begin{equation}\label{RHD2021ClLM cont eq expansion order 1 integrated} n_{0s}v_{sz1}=Un_{s1}. \end{equation}

Next, we consider the Poisson equation
\begin{equation}\label{RHD2021ClLM Poisson equation I order} n_{e1}-n_{i1}=0 \end{equation}
and
\begin{equation}\label{RHD2021ClLM Poisson equation II order} -\partial_{\xi}^{2}\varphi_{1}=4\pi (q_{e}n_{e2}+q_{i}n_{i2}). \end{equation}

Necessary relation between concentration, velocity field and electric field is found from the Euler equation
which is also considered in the first and second orders of expansion
\begin{equation}\label{RHD2021ClLM Euler equation I order} -Un_{0s}\partial_{\xi}v_{sz1}+u_{sp}^{2}\partial_{\xi}n_{s1}
=-\frac{q_{s}}{m_{s}}n_{0s}\biggl(\frac{\Gamma_{0s}}{n_{0s}}-\frac{u_{st}^{2}}{c^{2}}\biggr)\partial_{\xi}\varphi_{1}, \end{equation}
and
$$-Un_{0s}\partial_{\xi}v_{sz2}+n_{0s}\partial_{\tau}v_{sz1}+u_{sp}^{2}\partial_{\xi}n_{s2}+\frac{u_{sp}^{2}}{3\gamma_{Fs}^{2}n_{0s}}\partial_{\xi}n_{s1}^{2}$$
\begin{equation}\label{RHD2021ClLM Euler equation II order} =-\frac{q_{s}}{m_{s}}\biggl(\Gamma_{s1}-\frac{u_{st}^{2}}{c^{2}}n_{s1} \biggr)\partial_{\xi}\varphi_{1}.\end{equation}
The Euler equation obtained in the first order (\ref{RHD2021ClLM Euler equation I order}) can be integrated.
So, necessary relation between $v_{sz1}$, $n_{s1}$, and $\varphi_{1}$ is found.
Relation between $v_{sz2}$, $n_{s2}$, $v_{sz1}$, and $\varphi_{1}$ presented by equation (\ref{RHD2021ClLM Euler equation II order})
includes the first order perturbation of the relativistic hydrodynamic gamma function $\Gamma_{s1}$.
Therefore, we consider equation (\ref{RHD2021ClLM eq for Gamma}) in the lowest order of expansion
\begin{equation}\label{RHD2021ClLM Gamma eq I order}
-U\partial_{\xi}\Gamma_{s1}+\Gamma_{0s}\partial_{\xi}v_{sz1}+\partial_{\xi}t_{sz1}=0. \end{equation}
Next, we also need equation for the first order perturbation of the flux of the relativistic hydrodynamic gamma function $t_{sz1}$:
$$U\partial_{\xi}t_{sz1}-u_{st}^{2}\partial_{\xi}n_{s1}+\frac{\Gamma_{0s}}{n_{0s}}u_{sp}^{2}\partial_{\xi}n_{s1}
+\frac{q_{s}}{m_{s}}\Gamma_{0s}
\biggl(\frac{\Gamma_{0s}}{n_{0s}}-\frac{u_{st}^{2}}{c^{2}}\biggr)\partial_{\xi}\varphi_{1}$$

\begin{equation}\label{RHD2021ClLM t evol eq I order} =\frac{q_{s}}{m_{s}}n_{0s} \biggl(1-\frac{5u_{sp}^{2}}{c^{2}}+\frac{10}{3}\frac{u_{Ms}^{4}}{c^{4}}\biggr)\partial_{\xi}\varphi_{1}, \end{equation}
where we introduce the characteristic velocity for function $M_{0s}$ as follows $u_{Ms}^{4}\equiv M_{0s}/n_{0s}$.
Let us to point out that equations (\ref{RHD2021ClLM Gamma eq I order}) and (\ref{RHD2021ClLM t evol eq I order}) are used in the second order.

In the first order we find the following expressions for the concentration as the function of the potential of the electric field
\begin{equation}\label{RHD2021ClLM n s1 via phi}
n_{s1}=\frac{q_{s}}{m_{s}}\biggl(\frac{\Gamma_{0s}}{n_{0s}}-\frac{u_{st}^{2}}{c^{2}}\biggr)
\frac{n_{0s}}{U^{2}-u_{sp}^{2}}\varphi_{1}. \end{equation}
Let us repeat that $\frac{\Gamma_{0s}}{n_{0s}}-\frac{u_{st}^{2}}{c^{2}}=\frac{1}{\gamma_{Fs}}$.
We substitute expression (\ref{RHD2021ClLM n s1 via phi}) in the Poisson equation (\ref{RHD2021ClLM Poisson equation I order})
and find equation for the velocity of perturbation $U$:
\begin{equation}\label{RHD2021ClLM eq for U}
\frac{1}{m_{e}}
\frac{\frac{\Gamma_{0e}}{n_{0e}}-\frac{u_{et}^{2}}{c^{2}}}{U^{2}-u_{ep}^{2}}
+\frac{1}{m_{i}}
\frac{\frac{\Gamma_{0i}}{n_{0i}}-\frac{u_{it}^{2}}{c^{2}}}{U^{2}-u_{ip}^{2}}=0.\end{equation}
Expressions $\frac{\Gamma_{0s}}{n_{0s}}-\frac{u_{st}^{2}}{c^{2}}>0$ are positive.
Hence equation (\ref{RHD2021ClLM eq for U}) has solution under condition $u_{ip}^{2}< U^{2}< u_{ep}^{2}$.
Moreover, it is well-known
that stable ion-acoustic waves exist at more strict condition
$u_{ip}^{2}\ll U^{2}\ll u_{ep}^{2}$.
It allows us to get corresponding approximate solution of equation (\ref{RHD2021ClLM eq for U})
\begin{equation}\label{RHD2021ClLM U in I order} U^{2}=\frac{m_{e}}{m_{i}}\frac{\gamma_{Fe}}{\gamma_{Fi}}u_{ep}^{2}. \end{equation}

The second order leads to the nonlinear equation for the electric potential
\begin{widetext}
$$\partial_{\xi}^{3}\varphi_{1}
+\sum_{s=e,i}\frac{U\omega_{Ls}^{2}}{\gamma_{Fs}(U^{2}-u_{sp}^{2})^{2}}
\partial_{\tau}\varphi_{1}
+\sum_{s=e,i}\frac{q_{s}}{m_{s}}
\frac{2(U^{2}+\frac{u_{sp}^{2}}{3\gamma_{Fs}^{2}})\omega_{Ls}^{2}}{\gamma_{Fs}^{2}(U^{2}-u_{sp}^{2})^{3}}\varphi_{1}\partial_{\xi}\varphi_{1}$$
$$+\sum_{s=e,i}\frac{q_{s}}{m_{s}}\frac{\omega_{Ls}^{2}}{\gamma_{Fs}(U^{2}-u_{sp}^{2})^{2}}
\biggl[\frac{\Gamma_{0s}}{n_{0s}}U\biggl(1-\frac{u_{sp}^{2}}{U^{2}}\biggr)
+\frac{1}{U}\biggl(\frac{v_{Fs}^{2}}{3\gamma_{Fs}}-u_{st}^{2}\frac{U^{2}}{c^{2}}\biggr)\biggr]\varphi_{1}\partial_{\xi}\varphi_{1}$$
\begin{equation}\label{RHD2021ClLM KdV simm}
+\sum_{s=e,i}\frac{q_{s}}{m_{s}}\frac{\omega_{Ls}^{2}}{U^{2}-u_{sp}^{2}}
\biggl(1-5\frac{p_{0s}}{n_{0s}c^{2}}-\frac{\Gamma_{0s}}{\gamma_{Fs}n_{0s}}+\frac{10}{3}\frac{u_{sM}^{4}}{c^{4}}\biggr)
\varphi_{1}\partial_{\xi}\varphi_{1}=0. \end{equation}
Equation (\ref{RHD2021ClLM KdV simm}) is given in the general form
which is symmetric relatively all species.
Let us include condition $u_{ip}^{2}\ll U^{2}\ll u_{ep}^{2}$
in equation (\ref{RHD2021ClLM KdV simm}).
So it transforms in the following form:
$$\partial_{\xi}^{3}\varphi_{1}
+\biggl(\frac{U\omega_{Le}^{2}}{\gamma_{Fe}u_{ep}^{4}}
+\frac{\omega_{Li}^{2}}{\gamma_{Fi}U^{3}} \biggr)
\partial_{\tau}\varphi_{1}
+\biggl(-\frac{q_{e}}{m_{e}}
\frac{2(U^{2}+\frac{u_{ep}^{2}}{3\gamma_{Fe}^{2}})\omega_{Le}^{2}}{\gamma_{Fe}^{2}u_{ep}^{6}}
+\frac{q_{i}}{m_{i}}
\frac{2(U^{2}+\frac{u_{ip}^{2}}{3\gamma_{Fi}^{2}})\omega_{Li}^{2}}{\gamma_{Fi}^{2}U^{6}} \biggr)
\varphi_{1}\partial_{\xi}\varphi_{1}$$

$$+\Biggl(\frac{q_{e}}{m_{e}}\frac{\omega_{Le}^{2}U}{\gamma_{Fe}u_{ep}^{4}}
\biggl[\frac{\Gamma_{0e}}{n_{0e}}\biggl(1-\frac{u_{ep}^{2}}{U^{2}}\biggr)
+\biggl(\frac{v_{Fe}^{2}}{3\gamma_{Fe}U^{2}}-\frac{u_{et}^{2}}{c^{2}}\biggr)\biggr]
+\frac{q_{i}}{m_{i}}\frac{\omega_{Li}^{2}}{\gamma_{Fi}U^{3}}
\biggl[\frac{\Gamma_{0i}}{n_{0i}}\biggl(1-\frac{u_{ip}^{2}}{U^{2}}\biggr)
+\biggl(\frac{v_{Fi}^{2}}{3\gamma_{Fi}U^{2}}-\frac{u_{it}^{2}}{c^{2}}\biggr)\biggr]\Biggr)\varphi_{1}\partial_{\xi}\varphi_{1}$$

\begin{equation}\label{RHD2021ClLM KdV ui U ue}
+\biggl(-\frac{q_{e}}{m_{e}}\frac{\omega_{Le}^{2}}{u_{ep}^{2}}
\biggl(1-5\frac{p_{0e}}{n_{0e}c^{2}}-\frac{\Gamma_{0e}}{\gamma_{Fe}n_{0e}}+\frac{10}{3}\frac{u_{eM}^{4}}{c^{4}}\biggr)
+\frac{q_{i}}{m_{i}}\frac{\omega_{Li}^{2}}{U^{2}}
\biggl(1-5\frac{p_{0i}}{n_{0i}c^{2}}-\frac{\Gamma_{0i}}{\gamma_{Fi}n_{0i}}+\frac{10}{3}\frac{u_{iM}^{4}}{c^{4}}\biggr)\biggr)
\varphi_{1}\partial_{\xi}\varphi_{1}=0,
\end{equation}
\end{widetext}
where we use $u_{ep}\sim c$, hence $U\ll c$.
Since we have $u_{ip}^{2}\ll U^{2}$
we can expect that $u_{it}^{2}\ll U^{2}$ and $u_{iM}^{4}\ll c^{4}$.

Condition $u_{ip}^{2}\ll U^{2}\ll u_{ep}^{2}$ together with assumption $u_{ep}\sim c$
($u_{ep}< c$, but they have same order)
leads to $u_{ip}^{2}\ll U^{2}\ll c$.
It allows us assume $\gamma_{Fi}\approx1$.
Moreover, the contribution of the relativistic effects in the properties of ions can be dropped.
Let us consider the second term in equation (\ref{RHD2021ClLM KdV ui U ue}).
It is proportional to $\frac{U\omega_{Le}^{2}}{\gamma_{Fe}u_{ep}^{4}}
+\frac{\omega_{Li}^{2}}{U^{3}}$,
where we included $\gamma_{Fi}=1$.
We have $\omega_{Le}^{2}\gg \omega_{Li}^{2}$,
but the contribution of $\omega_{Le}^{2}$ is reduced by $\gamma_{Fe}u_{ep}^{4}$,
which is large in compare with $U^{4}$:
$\sqrt{\gamma_{Fe}}u_{ep}^{2} > u_{ep}^{2}\gg U^{2}$.
Hence, both terms are comparable under these conditions.

Next, we consider the coefficient in the third term in equation (\ref{RHD2021ClLM KdV ui U ue}).
We have $\gamma_{Fe}^{2}>1$, but we can also have $\gamma_{Fe}^{2}\gg1$.
Therefore, parameter $u_{ep}^{2}/3\gamma_{Fe}^{2}$ is reduced in compare with $u_{ep}^{2}$,
consequently the parameter $u_{ep}^{2}/3\gamma_{Fe}^{2}$ is comparable with $U^{2}$.
Here we have stronger reduction of the electron Langmuir frequency square in compare with the similar reduction of the electron Langmuir frequency square second term of equation (\ref{RHD2021ClLM KdV ui U ue}).
We have the following construction $[(U^{2}+\frac{u_{ep}^{2}}{3\gamma_{Fe}^{2}})/u_{ep}^{2}]\omega_{Le}^{2}/\gamma_{Fe}^{2}u_{ep}^{4}$,
where $[(U^{2}+\frac{u_{ep}^{2}}{3\gamma_{Fe}^{2}})/u_{ep}^{2}]\ll 1$
and $\gamma_{Fe}^{2}u_{ep}^{4}> u_{ep}^{4}\gg U^{4}$.
We estimate the ion contribution in the coefficient in the third term in equation (\ref{RHD2021ClLM KdV ui U ue}),
where we find $(U^{2}+\frac{u_{ip}^{2}}{3})\omega_{Li}^{2}/U^{6}\approx \omega_{Li}^{2}/U^{4}$.
It shows that contributions of electrons and ions can be comparable
or one of them can dominate.
If both parts of coefficient in the second term in equation (\ref{RHD2021ClLM KdV ui U ue}) are comparable
we can drop the contribution of electrons in the coefficient in the third term in equation (\ref{RHD2021ClLM KdV ui U ue}).
In the opposite limit, if both parts of the coefficient in the third term in equation (\ref{RHD2021ClLM KdV ui U ue}) are comparable
we can drop the contribution of ions in coefficient in the second term in equation (\ref{RHD2021ClLM KdV ui U ue}).
In general case, we keep all of them.

We consider the electron contribution in the coefficient of the fourth term in equation (\ref{RHD2021ClLM KdV ui U ue}),
where we drop $1$ in compare with $u_{ep}^{2}/U^{2}$,
we can drop $\frac{u_{et}^{2}}{c^{2}}<1$ in compare with $\frac{v_{Fe}^{2}}{3\gamma_{Fe}U^{2}}\gg 1$ (if relativistic effects are relatively small),
but we have $\frac{v_{Fe}^{2}}{3\gamma_{Fe}U^{2}}\geq 1$ (comparable with $\frac{u_{et}^{2}}{c^{2}}\sim1$)
for the strong relativistic effects $\gamma_{Fe}\gg 1$.
We also have a product of small $\frac{\Gamma_{0e}}{n_{0e}}$ and large $\frac{u_{ep}^{2}}{U^{2}}$ parameters $\frac{\Gamma_{0e}}{n_{0e}}\frac{u_{ep}^{2}}{U^{2}}$
which can be above or below of $1$.

We consider the contribution of ions in the coefficient of the fourth term in equation (\ref{RHD2021ClLM KdV ui U ue}).
We include $u_{ip}^{2}/U^{2}\ll1$
We also use $\Gamma_{0i}\approx n_{0i}$.
We see the combination of two terms $\frac{v_{Fi}^{2}}{3U^{2}}\ll 1$ and $\frac{u_{it}^{2}}{c^{2}}\ll1$,
but they can be comparable to each other.
So, we have $1+(\frac{v_{Fi}^{2}}{3U^{2}}-\frac{u_{it}^{2}}{c^{2}})$,
where we can drop two last terms in compare with $1$.
So, we have simplification of the contribution of ions down to $\frac{q_{i}}{m_{i}}\frac{\omega_{Li}^{2}}{U^{3}}$.
In general case it can be comparable with the contribution of electrons.

Finally, we present an analysis of the last term in equation (\ref{RHD2021ClLM KdV ui U ue}).
No further simplification is found for the part presenting electrons.
However, strong simplification is found for the ions,
where
$u_{iM}^{4}\ll c^{4}$,
$p_{0i}/n_{0i}c^{2}\sim v_{Fi}^{2}/c^{2}\ll 1$.
Hence, two terms are found $1-\frac{\Gamma_{0i}}{n_{0i}}=0$
which give zero combination
since $\Gamma_{0i}\approx n_{0i}$.
So, there is no contribution of ions in the last term.

After described modifications we find simplification of equation (\ref{RHD2021ClLM KdV ui U ue}):
\begin{widetext}

$$\partial_{\xi}^{3}\varphi_{1}
+\biggl(\frac{U\omega_{Le}^{2}}{\gamma_{Fe}u_{ep}^{4}}
+\frac{\omega_{Li}^{2}}{U^{3}} \biggr)
\partial_{\tau}\varphi_{1}
+\biggl(-\frac{q_{e}}{m_{e}}
\frac{2(U^{2}+\frac{u_{ep}^{2}}{3\gamma_{Fe}^{2}})\omega_{Le}^{2}}{\gamma_{Fe}^{2}u_{ep}^{6}}
+\frac{q_{i}}{m_{i}}
\frac{2\omega_{Li}^{2}}{U^{4}} \biggr)
\varphi_{1}\partial_{\xi}\varphi_{1}$$

$$+\Biggl(\frac{q_{e}}{m_{e}}\frac{\omega_{Le}^{2}U}{\gamma_{Fe}u_{ep}^{4}}
\biggl[-\frac{\Gamma_{0e}}{n_{0e}}\frac{u_{ep}^{2}}{U^{2}}
+\biggl(\frac{v_{Fe}^{2}}{3\gamma_{Fe}U^{2}}-\frac{u_{et}^{2}}{c^{2}}\biggr)\biggr]
+\frac{q_{i}}{m_{i}}\frac{\omega_{Li}^{2}}{U^{3}}
\Biggr)\varphi_{1}\partial_{\xi}\varphi_{1}$$

\begin{equation}\label{RHD2021ClLM KdV ui U ue 2}
-\frac{q_{e}}{m_{e}}\frac{\omega_{Le}^{2}}{u_{ep}^{2}}
\biggl(1-5\frac{p_{0e}}{n_{0e}c^{2}}-\frac{\Gamma_{0e}}{\gamma_{Fe}n_{0e}}+\frac{10}{3}\frac{u_{eM}^{4}}{c^{4}}\biggr)
\varphi_{1}\partial_{\xi}\varphi_{1}=0.
\end{equation}
\end{widetext}
The relativistic effects contribute in the properties of the ion-acoustic soliton described by equation (\ref{RHD2021ClLM KdV ui U ue 2})
via the electrons.
Main indicator of these effects is the gamma factor on the Fermi velocity $\gamma_{Fe}$.
It is located in the denumerators.
Hence, it reduces the contribution of electrons in compare with the ions.


\section{Conclusion}

General derivation of the hydrodynamic model for the relativistically hot plasmas has been presented in earlier papers
\cite{Andreev 2021 05}, \cite{Andreev 2021 09}.
This hydrodynamic model is based on the dynamics of
four material fields: the concentration and the velocity field
\emph{and} the average reverse relativistic $\gamma$ factor and the flux of the reverse relativistic $\gamma$ factor.
Here we have presented further generalization of this model for the degenerate species.
So, we consider temperatures below the Fermi temperature of the chosen species.
Moreover, the concentration of the species is large enough
so the Fermi velocity getting close to the speed of light.
Necessary equations of state for the flux of the particle current $\tilde{p}$,
the flux of the average reverse gamma factor $\tilde{t}$,
and function $M$,
which is the flux of flux of function $\tilde{p}$,
have been obtained in the paper.
Moreover, the equilibrium average reverse gamma factor $\Gamma_{0}$ has been calculated for degenerate fluid as well.

Major application of the developed model has been made for the small amplitude ion-acoustic soliton.
However, in order to illustrate the role of the relativistic effects on the simple example,
we have considered the spectrum of the Langmuir waves.
Moreover, Langmuir waves are considered in two ways.
First, the suggested model has been applied to find the spectrum of the Langmuir waves.
Second, the relativistic Vlasov kinetic equation has been used to consider the same problem
in order to estimate the accuracy of the suggested model.
The properties of the relativistic ion-acoustic solitons are studied analytically in terms of the suggested model.

\section{Acknowledgements}

Work is supported by the Russian Foundation for Basic Research
(grant no. 20-02-00476).

\section{DATA AVAILABILITY}

Data sharing is not applicable to this article as no new data were
created or analyzed in this study, which is a purely theoretical one.

\end{document}